\documentclass[pra,aps,twocolumn]{revtex4}

\usepackage{graphicx}

\begin{document}

\title{Mean-field analysis of collapsing and exploding Bose-Einstein
condensates}

\author{Hiroki Saito}
\author{Masahito Ueda}
\affiliation{Department of Physics, Tokyo Institute of Technology,
Tokyo 152-8551, Japan}

\date{\today}

\begin{abstract}
The dynamics of collapsing and exploding trapped Bose-Einstein condensates
caused by a sudden switch of interactions from repulsive to attractive are
studied by numerically integrating the Gross-Pitaevskii equation with
atomic loss for an axially symmetric trap.
We investigate the decay rate of condensates and the phenomena of bursts
and jets of atoms, and compare our results with those of the experiments
performed by E. A. Donley {\it et al.} [Nature {\bf 412}, 295 (2001)].
Our study suggests that the condensate decay and the burst production is
due to local intermittent implosions in the condensate, and that atomic
clouds of bursts and jets are coherent.
We also predict nonlinear pattern formation caused by the density
instability of attractive condensates.
\end{abstract}

\maketitle

\section{Introduction}

Bose-Einstein condensates (BECs) of trapped atomic vapor have been
realized in several atomic
species~\cite{Anderson,Davis,Bradley,Fried,Robert,Modugno}.
One of the remarkable features of these systems is that both
strength~\cite{Inouye} and sign~\cite{Cornish,Roberts01,Donley} of
interactions between atoms can be tuned by adjusting an external magnetic
field near a Feshbach resonance~\cite{Feshbach}.
This has opened up the possibility to study collapsing and exploding BECs
in a controllable manner, thereby offering a stringent test of the
Gross-Pitaevskii mean-field theory~\cite{GP}.

A trapped BEC with attractive interactions may be formed~\cite{Ruprecht}
when quantum pressure arising from Heisenberg's uncertain principle
counterbalances the attractive force between atoms.
This is possible when the parameter defined by $k \equiv N_0 |a| / d_0$ is
below a critical value $k_{\rm cr}$, where $N_0$ is the number of BEC
atoms, $a$ the s-wave scattering length, and $d_0$ the size of the
ground-state wave function of a harmonic trap.
When the parameter $k$ exceeds $k_{\rm cr}$, attractive force dominates
quantum pressure, causing BEC to collapse.
In the experiments performed by a Rice
group~\cite{Sackett,Sackett99,Gerton01}, BEC atoms are continuously
supplied from a supercooled thermal gas, and collapse and growth cycles of
BEC have been observed~\cite{Gerton01}.
In the experiments performed at JILA~\cite{Cornish,Roberts01,Donley}, in
contrast, $N_0$ is fixed and $a (< 0)$ is decreased by using the Feshbach
resonance so as to meet the condition $k > k_{\rm cr}$.
This technique enabled them to determine the value of $k_{\rm
cr}$~\cite{Roberts01}, and to observe exploding atomic ejection from
collapsing BEC~\cite{Cornish,Donley}, a phenomenon called ``Bosenova''
whose origin is currently under
controversy~\cite{Kagan98,SaitoL,SaitoA,Duine}.

The Gross-Pitaevskii (GP) equation~\cite{GP} has widely been used to study
mean-field properties of BECs, offering fairly good quantitative account
of a rich variety of experiments for repulsive BECs~\cite{Dalfovo}.
In the case of the attractive BECs, however, it is by no means clear to
what extent mean-field theory is valid, for the attractive interaction
might enhance many-body quantum correlations.
In fact, a measured value of $k_{\rm cr}$~\cite{Roberts01} is
significantly smaller than that predicted by a mean-field
theory~\cite{Ruprecht,Gammal}, the origin of the discrepancy being not yet
understood.

The recent quantitative measurements on collapsing and exploding BEC
reported in Ref.~\cite{Donley} have motivated us to investigate to what
extent mean-field theory can explain the experimental observations.
This is the main purpose of this paper.
By numerically integrating the GP equation with atomic loss, we show that
the phenomena reported in Ref.~\cite{Donley}, such as decay of
condensates, ejection and refocus of atomic bursts, and jet formation, are
reproduced by our numerical simulations.
In addition, we predict that various patterns in atomic density are formed
in the course of collapse.

This paper is organized as follows.
Section \ref{s:GP} briefly reviews the GP equation with atomic loss due to 
inelastic collisions and describes the method of analysis of our
simulations.
Section \ref{s:numerical} reports our results of numerical simulations
using the GP equation, and compares them with the experimental data of
Ref.~\cite{Donley}.
Section \ref{s:conclusion} provides the summary of this paper.

\section{The Gross-Pitaevskii equation with atomic loss and methods of
analysis}
\label{s:GP}

The GP equation describes unitary time evolution of a macroscopic ``wave
function'' $\psi$, and conserves the total number of atoms $\int |\psi|^2
d{\bf r}$.
In reality, however, atoms are lost from the trap due to the two-body
dipolar and three-body recombination losses.
These effects may be taken into account by incorporating in the GP
equation the imaginary terms describing these inelastic
processes~\cite{Kagan98}:
\begin{eqnarray} \label{GP}
i \hbar \frac{\partial}{\partial t} \psi & = & -\frac{\hbar^2}{2m} 
\nabla^2 \psi + V_{\rm trap}({\bf r}) \psi + \frac{4 \pi \hbar^2 a}{m}
|\psi|^2 \psi \nonumber \\
& & - \frac{i \hbar}{2} \left( K_2 |\psi|^2 + K_3 |\psi|^4 \right) \psi,
\end{eqnarray}
where $V_{\rm trap}$ is the trapping potential, and $K_2$ and $K_3$ denote
two-body dipolar and three-body recombination loss-rate coefficients,
respectively.
The imaginary terms in Eq.~(\ref{GP}) are phenomenologically introduced in
order to take account of the fact that the two-body and three-body losses
are proportional to the square and cube of the atomic density:
\begin{equation}
\frac{\partial}{\partial t} \int |\psi|^2 d{\bf r} = -\int (K_2 |\psi|^4 +
K_3 |\psi|^6) d{\bf r}. 
\end{equation}
We assume that the atoms and molecules produced by inelastic collisions
escape from the trap without affecting the condensate.
The constants $K_2$ and $K_3$ include Bose statistical factors $1 / 2!$
and $1 / 3!$, respectively, which are needed for BEC~\cite{factor}, and
describe the loss rate per atom.

In the situations we consider the two-body loss can be ignored.
The two-body and three-body loss rates are given by $R_2 \equiv
K_2 |\psi|^2$ and $R_3 \equiv K_3 |\psi|^4$, and their ratio by $R_{23}
\equiv R_2 / R_3 = K_2 / (K_3 |\psi|^2)$.
They have the relation $R_2 R_{23} = K_2^2 / K_3$, which is always
$\lesssim 0.1 \; {\rm s}^{-1}$ for ${}^{85}{\rm Rb}$~\cite{Roberts00} and
${}^7{\rm Li}$~\cite{Gerton}.
When the density is low and $R_2 \sim R_3$, $R_2 \lesssim 0.1 \; {\rm
s}^{-1}$.
Then both two-body and three-body losses can be ignored since we shall
consider the time scale of $\lesssim 10$ ms.
When the density is high, the three-body loss dominates the two-body loss,
and thus the two-body loss becomes unimportant.
In our numerical simulations we shall therefore ignore the two-body loss
term in Eq.~(\ref{GP}).
All results presented below are not affected if the two-body loss is taken
into account.

The three-body loss, in contrast, plays a crucial role in determining
the collapsing dynamics of BEC.
When implosion occurs, the atomic density becomes extremely high, and so
does the three-body recombination rate, until it stops the growth of the
density.
The maximum density in the process of implosion is determined by $K_3$ and
$a$~\cite{Kagan98,SaitoL,SaitoA}.
The values of $K_3$ far from the Feshbach resonance have been measured for
${}^{87}{\rm Rb}$~\cite{Burt}, ${}^{23}{\rm Na}$~\cite{Kurn}, and
${}^7{\rm Li}$~\cite{Gerton}, and they agree with theoretical
predictions~\cite{Moer,Fedichev,Esry} within a factor of ten.
Near the Feshbach resonance, however, complicated behaviors of $K_3$ are
predicted~\cite{Braaten}, with no precise experimental data
available~\cite{Roberts00}.

We performed numerical integration of the GP equation (\ref{GP}) using a
finite-difference method with the Crank-Nicholson scheme~\cite{Ruprecht}.
Since the peak density changes drastically during
implosion~\cite{SaitoL,SaitoA}, we very carefully controlled the time step
to avoid error propagation.
Initially we prepared the ground-state wave function by the method in
Ref.~\cite{Edwards} for an initial s-wave scattering length $a_{\rm init}$
and an initial number of BEC atoms $N_0$.
At $t = 0$ the interaction is suddenly switched from $a_{\rm init} \geq 0$
to $a_{\rm collapse} < 0$, inducing collapse of the condensate.
We use the parameters of ${}^{85}{\rm Rb}$ and assume the same trap
geometry (i.e. radial frequency $\omega_\perp / 2\pi = 17.5$ Hz and axial
frequency $\omega_z / 2\pi = 6.8$ Hz) as that used in Ref.~\cite{Donley}.

In Ref.~\cite{Donley}, atoms after the collapse are classified into three
parts: remnant, burst, and missing atoms.
The remnant BEC is a dense atomic cloud peaking around the center of the
trap, and the burst is a dilute one that spreads broadly around the
remnant BEC.
In our simulations, we identify the remnant, burst, and missing atoms as
follows.
The one-dimensional density distributions defined by $\rho_{\rm axial}(z)
\equiv \int |\psi|^2 dx dy$ and $\rho_{\rm radial}(r) \equiv \int |\psi|^2
dz$ show bimodal structures, one peaking at the center and the other
spreading around it (an example is shown in Fig.~\ref{f:burst}b).
We identify the former as the remnant and the latter as the burst, and
determine the axial and radial coordinates, $z_b$ and $r_b$, of the
boundaries between the remnant BEC and the burst.
We then calculate the number of atoms in the remnant BEC as
\begin{equation}
N_{\rm remnant} \equiv \int_0^{r_b} 2 \pi r dr \int_{-z_b}^{z_b} dz
|\psi|^2.
\end{equation}
The burst atoms are defined by the ones outside the boundary.
Because of ambiguities in defining the coordinates of the boundaries,
$N_{\rm remnant}$ and $N_{\rm burst}$ defined this way have uncertainties
of about $\pm 0.05 N_0$, while the total number of atoms in the trap
$N_{\rm tot} \equiv N_{\rm remnant} + N_{\rm burst}$ is well defined.
We define the number of missing atoms as $N_{\rm missing} \equiv N_0 -
N_{\rm tot}$.

\section{Results of numerical simulations}
\label{s:numerical}

\subsection{Decay of the condensate and local intermittent implosions}

\begin{figure}[tb]
\includegraphics[width=8.6cm]{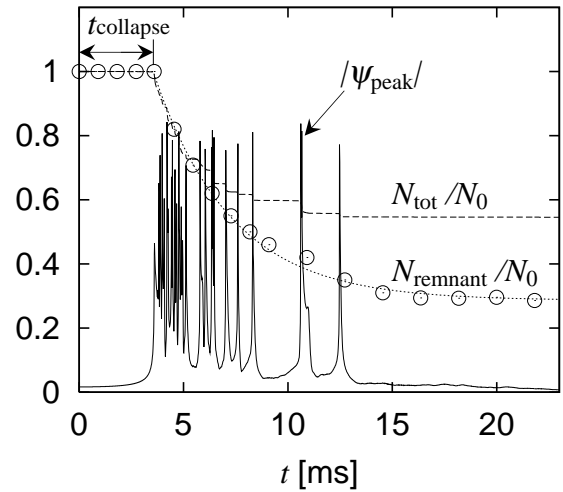}
\caption{
Time evolutions of the peak height of the wave function $|\psi_{\rm
peak}|$ (solid curve in arbitrary scale), the fraction $N_{\rm tot} / N_0$
of the total number of atoms remaining in the trap (dashed curve), and the
fraction $N_{\rm remnant} / N_0$ of the remnant BEC atoms (circles).
The dotted curve is the best fit of the circles to
Eq.~(\protect\ref{exp}) with $t_{\rm collapse} = 3.6$ ms and $\tau_{\rm
decay} = 3.7$ ms.
At $t = 0$ the s-wave scattering length is changed from $a_{\rm init} = 7
a_0$ to $a_{\rm collapse} = -30 a_0$, where $a_0$ is the Bohr radius.
The initial number of BEC atoms is $N_0 = 15000$, and the three-body
recombination loss-rate coefficient is $K_3 = 2 \times 10^{-28} {\rm cm}^6
/ {\rm s}$.
}
\label{f:decay}
\end{figure}

We first study the situation in which the scattering length is switched
from $a_{\rm init} = 7 a_0$ to $a_{\rm collapse} = -30 a_0$ at $t = 0$,
where $a_0$ is the Bohr radius.
Figure \ref{f:decay} shows time evolutions of the peak height $|\psi_{\rm
peak}|$ of the wave function, the total number of atoms in the trap
$N_{\rm tot}$, and the number of remnant BEC atoms $N_{\rm remnant}$,
where we assume $K_3 = 2 \times 10^{-28} {\rm cm}^6 / {\rm s}$.

Our numerical simulations show that the condensate first contracts slowly
with its peak height $|\psi_{\rm peak}|$ gradually increasing.
The total number of atoms remains constant during this process, since the
recombination loss is negligible at such low densities.
At $t \simeq 3.6$ ms, implosion suddenly occurs in a very localized region
(the size of the density spike is $\sim 0.1 \; \mu{\rm m}$ while the size
of the BEC cloud is several micrometers) for a very short period of time
($\sim 0.1$ ms).
Furthermore the implosion occurs not just once but many times
intermittently for about 10 ms~\cite{SaitoL,SaitoA}.
The three-body recombination loss prominently occurs during the implosion,
since the atomic density becomes extremely high.
Several tens of atoms are lost in each intermittent implosion, resulting
in a step-wise decrease of $N_{\rm tot}$.
The decay of $N_{\rm remnant}$ is due to both three-body recombination and
atomic burst ejection.
We note that the behavior of $N_{\rm remnant}$, shown as the plots in
Fig.~\ref{f:decay}, is very similar to the experimental result (Fig.~1b of
Ref.~\cite{Donley}).

\begin{figure}[tb]
\includegraphics[width=8.6cm]{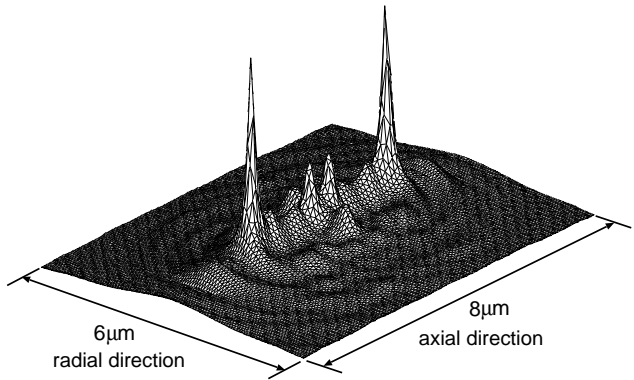}
\caption{
A snapshot of the column density of the imploding BEC taken at 2.5 ms
after the s-wave scattering length is changed from $a_{\rm init} = 0$ to
$a_{\rm collapse} = -30 a_0$.
The initial number of BEC atoms is $N_0 = 15000$, and the loss-rate
coefficient is $K_3 = 2 \times 10^{-28} {\rm cm}^6 / {\rm s}$.
}
\label{f:snapshot}
\end{figure}

The implosions occur not only at the center of the trap but also at other
locations on the trap axis, and more than one density spike is often seen
simultaneously.
A snapshot of the imploding process with $a_{\rm init} = 0$, $a_{\rm
collapse} = -30 a_0$, and $N_0 = 15000$ is displayed in
Fig.~\ref{f:snapshot}, where the image is taken 2.5 ms after the switch of
interactions.
Two large spikes are seen on the trap axis.
In an isotropic trap, on the other hand, implosions always occur one by
one at the center of the trap.

\subsection{Collapse and decay times}

\begin{figure}[tb]
\includegraphics[width=8.6cm]{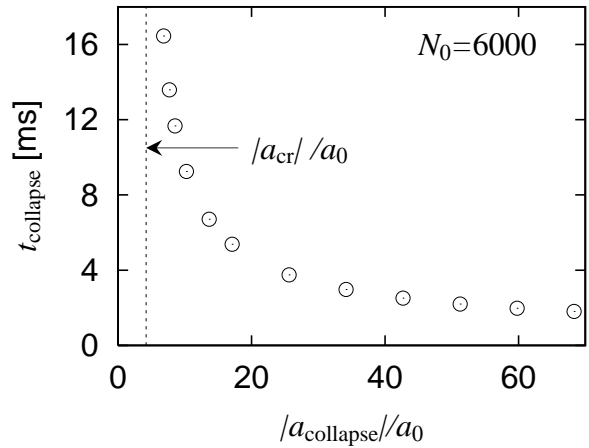}
\caption{
The collapse time $t_{\rm collapse}$ at which implosions begin as a
function of $|a_{\rm collapse}| / a_0$, where the initial number of BEC
atoms is $N_0 = 6000$.
At $t = 0$ the s-wave scattering length is switched from $a_{\rm init} =
0$ to $a_{\rm collapse} < 0$.
The dashed line indicates the critical ratio $|a_{\rm cr}| / a_0$ for $N_0
= 6000$, below which the condensate does not collapse.
}
\label{f:collapset}
\end{figure}

The total number of atoms of the system remains constant for some time
after the switch of interactions, and suddenly it begins to decay.
We call the time at which the sudden decay begins `collapse time' $t_{\rm
collapse}$ (see Fig.~\ref{f:decay}).
Since the collapse time is determined mainly by slow accumulation of atoms
towards the center of the trap, $t_{\rm collapse}$ only weakly depends on
the value of $K_3$.
In Fig.~\ref{f:decay}, $t_{\rm collapse} = 3.6$ ms, which agrees with the
experimental result of 3.7 ms~\cite{Donley}.
Figure~\ref{f:collapset} shows the dependence of $t_{\rm collapse}$ on
$|a_{\rm collapse}| / a_0$, where $a_{\rm init} = 0$ and $N_0 =
6000$~\cite{Elef}.
The plots are in good agreement with the experimental ones (Fig.~2 of
Ref.~\cite{Donley}).
For another set of parameters $a_{\rm init} = 89 a_0$ and $a_{\rm
collapse} = -15 a_0$, we obtain $t_{\rm collapse} = 15.9$ ms, which is
also consistent with the experimental finding~\cite{Donley}.
The collapse time increases for larger values of $a_{\rm init}$, since the
atomic cloud spreads more widely and therefore it takes longer time for
the cloud to get to the center of the trap.

Fitting
\begin{eqnarray} \label{exp}
N_{\rm remnant}(t) & = & [N_0 - N_{\rm remnant}(\infty)] \;
e^{-\frac{t - t_{\rm collapse}}{\tau_{\rm decay}}} \nonumber \\
& & + N_{\rm remnant}(\infty)
\end{eqnarray}
to the plots in Fig.~\ref{f:decay} (dotted curve), we obtain the decay
time constant $\tau_{\rm decay} \simeq 3.7$ ms.
For $a_{\rm init} = 0$, and $N_0 = 6000$ and $15000$, $\tau_{\rm decay}$
is almost the same as the above one, and for $a_{\rm collapse} = -60 a_0$
it becomes $\tau_{\rm decay} \simeq 3.1$ ms which agrees reasonably well
with the experimental finding of 2.8 ms~\cite{Donley}.

\subsection{Remnant, burst, and missing atoms}
\label{s:burst}

\begin{figure}[tb]
\includegraphics[width=8.6cm]{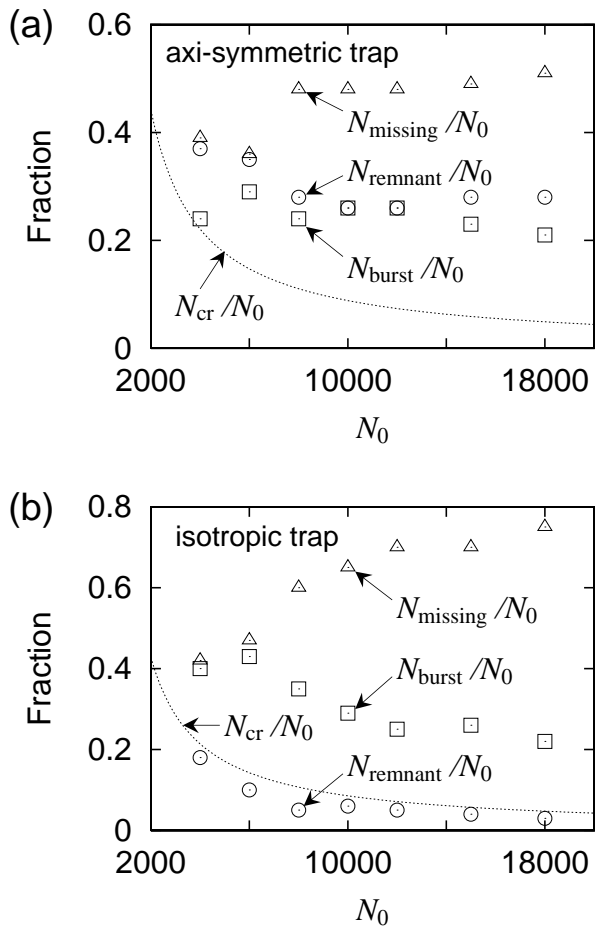}
\caption{
The fractions of the remnant $N_{\rm remnant} / N_0$ (circles), burst
$N_{\rm burst} / N_0$ (squares), and missing atoms $N_{\rm
missing} / N_0$ (triangles) after the collapse for (a) an
axi-symmetric trap and (b) an isotropic trap, where $N_0$ is the
initial number of BEC atoms.
The dotted curves show the fraction $N_{\rm cr} / N_0$ of the critical
number of atoms.
The s-wave scattering length is switched from $a_{\rm init} = 0$ to
$a_{\rm collapse} = -30 a_0$ with the loss-rate coefficient $K_3 = 2
\times 10^{-28} {\rm cm}^6 / {\rm s}$.
}
\label{f:fraction}
\end{figure}

Figure~\ref{f:fraction}a shows the fractions of remnant, burst, and
missing atoms after the implosion has finished for $a_{\rm init} = 0$ and
$a_{\rm collapse} = -30 a_0$.
We note that $N_{\rm remnant}$ is much larger than $N_{\rm cr}$ when $N_0$
is large, as observed experimentally~\cite{Donley}.
To put it differently, once the condensate expands following the collapse,
the critical density will not be reached even when $N_{\rm remnant} >
N_{\rm cr}$.
This is because the ratio $\omega_\perp / \omega_z$ is irrational and
therefore the axial and radial refocuses of the burst atoms do not occur
simultaneously, with no further implosions occurring.

This result presents a sharp contrast with that for an isotropic case
shown in Fig.~\ref{f:fraction}b, where the trap frequency is chosen to be
the geometric mean $(\omega_\perp^2 \omega_z)^{1/3}$.
The numbers of remnant atoms $N_{\rm remnant}$ are always below $N_{\rm
cr}$.
This is because all collapsing atoms go to the center of the trap
simultaneously, and therefore more implosions occur in an isotropic trap,
which increases $N_{\rm missing}$ and decreases $N_{\rm remnant}$.
Moreover, the implosions occur also when the burst atoms refocus, since
they concentrate at the center of the trap.
The data in Fig.~\ref{f:fraction}b are taken before the first refocus.
Nevertheless, we note that $N_{\rm remnant}$ is already below $N_{\rm
cr}$.

We also note that in the axi-symmetric trap the fractions $N_{\rm remnant}
/ N_0$, $N_{\rm burst} / N_0$, and $N_{\rm missing} / N_0$ are almost
independent of $N_0$, particularly for $N_0 > 6000$, which is consistent
with the experiments~\cite{Donley}.
This is a consequence of the fact that the number of implosions occurring
in the collapse is roughly proportional to $N_0$ and that the numbers of
the burst and missing atoms in each implosion are almost constant.

\subsection{Atomic bursts and ``jets''}

\begin{figure}[tb]
\includegraphics[width=8.6cm]{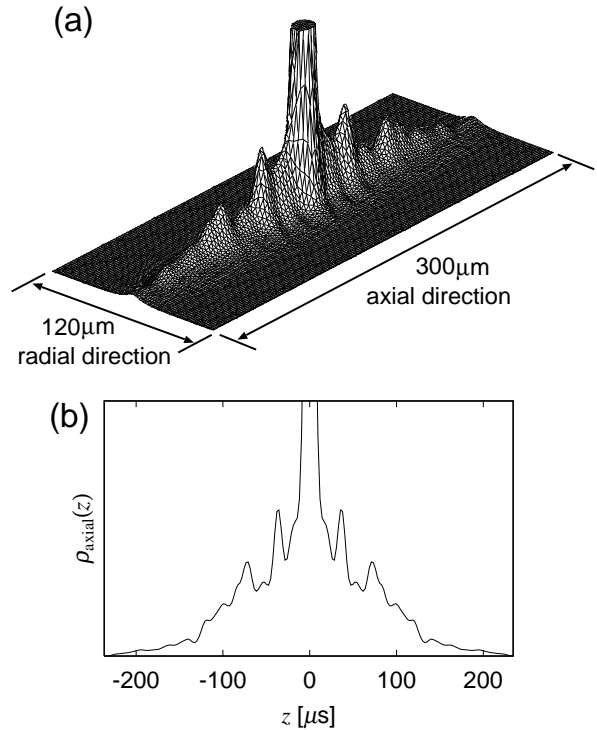}
\caption{
(a) The integrated column density $\rho(x, z) = \int |\psi|^2 dy$ seen
from the direction perpendicular to the trap axis and (b) the
one-dimensional density distribution $\rho_{\rm axial}(z) = \int |\psi|^2
dx dy$ along the axial direction.
The s-wave scattering length is switched from $a_{\rm init} = 0$ to
$a_{\rm collapse} = -30 a_0$ with $K_3 = 2 \times 10^{-28} {\rm cm}^6 /
{\rm s}$, where the images are taken at $t = 33.6$ ms.
The images are smoothed in accordance with the experimental resolution ($7
\; \mu{\rm m}$ FWHM), and the central peaks are truncated.
}
\label{f:burst}
\end{figure}

A burst atom cloud is usually too broadly spread and hence too dilute to
be seen.
However, at every $\pi / \omega_\perp$ (or $\pi / \omega_z$) the cloud
refocuses along the axial (or radial) direction and can be observed.
Figure~\ref{f:burst}a shows the column density seen from the direction
perpendicular to the trap axis and Fig.~\ref{f:burst}b shows the
one-dimensional distribution $\rho_{\rm axial}(z) \equiv \int |\psi|^2 dx
dy$, when the burst atoms focus along the trap axis (corresponding to
Fig.~3 of Ref.~\cite{Donley}).
The s-wave scattering length is switched from $a_{\rm init} = 0$ to
$a_{\rm collapse} = -30 a_0$ at $t = 0$ and the image is taken at $t =
33.6$ ms $\simeq t_{\rm collapse} + \pi / \omega_\perp$.
There are small peaks in the ridge of focus, which appear to correspond to
the shoulders of the central peak in Fig.~3c of Ref.~\cite{Donley}.
According to our theory the origin of the atomic burst is a release of
kinetic energy of local spikes in the atomic
density~\cite{Kagan98,SaitoL,SaitoA}.

As regards the burst energy, it is at present difficult to compare our
results with the experimental ones.
One reason is that although the burst energy depends not only on $a_{\rm
collapse}$ but also on $K_3$~\cite{SaitoA}, experimental values of $K_3$
as a function of $a_{\rm collapse}$ are not available.
The uncertainty in the boundaries between the remnant BEC and the burst
and that in Gaussian fitting of the burst profile (see
Fig.~\ref{f:burst}b) give rise to large errors in numerically determining
the burst ``temperature'', which also make quantitative comparison between
theory and experiment difficult.
More work (both theoretical and experimental) needs to be done in order to
clarify the situation.

In the experiments~\cite{Donley}, prolonged atomic clouds called ``jets''
are observed in the radial direction when the collapse is interrupted by
switching the interaction from attractive to repulsive.
The jets are distinguished from the bursts in that the energy of the
former is much lower than the latter and that the direction of the jets is
purely radial.
In Ref.~\cite{Donley}, the origin of the jets is considered to be the
highly anisotropic spikes in the atomic density.

\begin{figure}[tb]
\includegraphics[width=8.6cm]{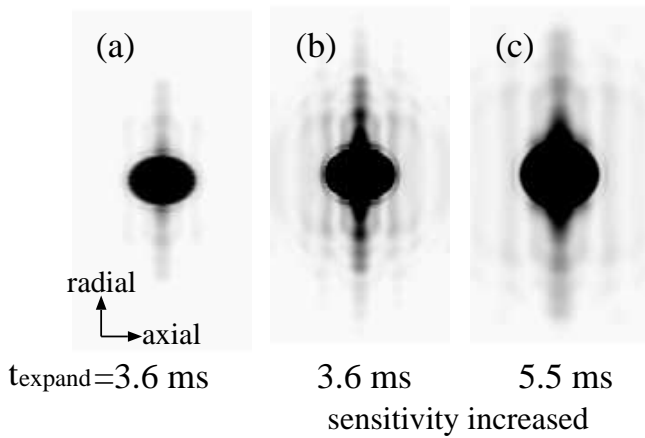}
\caption{
The integrated column densities $\rho(x, z) = \int |\psi|^2 dy$ of
condensates and jets seen from the radial direction.
The wave function in Fig.~\protect\ref{f:snapshot} is expanded by a switch
of the s-wave scattering length from $a_{\rm collapse} = -30 a_0$ to
$a_{\rm expand} = 100 a_0$.
The size of each image is $48 \times 90 \; \mu{\rm m}$.
(a) The image is taken with $t_{\rm expand} = 3.6$ ms. (b) The
sensitivity of imaging in (a) is increased. (c) The image is taken with
$t_{\rm expand} = 5.5$ ms and with the same sensitivity as (b).
}
\label{f:jets}
\end{figure}

We performed numerical simulations under situations similar to the
experimental ones.
The s-wave scattering length is switched from $a_{\rm init} = 0$ to
$a_{\rm collapse} = -30 a_0$ at $t = 0$, and changed to $a_{\rm expand} =
100 a_0$ at $t = 2.5$ ms (a snapshot before expansion is shown in
Fig.~\ref{f:snapshot}).
Figure~\ref{f:jets} shows the gray-scale images of the integrated column
density $\rho(x, z) = \int |\psi|^2 dy$ after a $t_{\rm expand} = 3.6$ ms
expansion in (a) and (b), and $t_{\rm expand} = 5.5$ ms in (c).
In the image (a), prolonged atomic clouds similar to the jets reported in
Fig.~5 of Ref.~\cite{Donley} appear in the radial direction (the main jet
at the center and two tiny jets on either side of it).
The images (b) and (c), where the sensitivity of the imaging is increased,
show that the jets are interference fringes.
The parallel fringe pattern is characteristic of the interference between
waves emanating from two point sources~\cite{Andrews}.
In fact, there are two spikes in the atomic density that play the role of
two point sources of matter waves in Fig.~\ref{f:snapshot}.
We find that the spacing between the fringes is proportional to $t_{\rm
expand}$.
This supports our interpretation, since the spacing between the fringes
from two point sources is known to be given by $ht_{\rm expand} /
md$~\cite{Andrews}, where $d$ is the distance between two point sources.
In the experimental images in Ref.~\cite{Donley} (particularly in
Figs.~5d-f), the jets seem to be ejected from the edge of the remnant
condensate, where the density is too low to form the spikes, which also
supports our interpretation that the jets are the consequence of
interference.
Thus, the experimental observations of jets indicate that atoms expanding
from spikes are coherent, suggesting that the burst atoms are also
coherent.

\subsection{Pattern formation}

We found in Ref.~\cite{SaitoL} that various patterns in the atomic density
are formed in the collapse processes caused by a sudden switch of
interactions from repulsive to attractive, the origin of the pattern
formation being attributed to the self-focusing effect of the attractive
systems.
\begin{figure}[tbp]
\includegraphics[width=8.6cm]{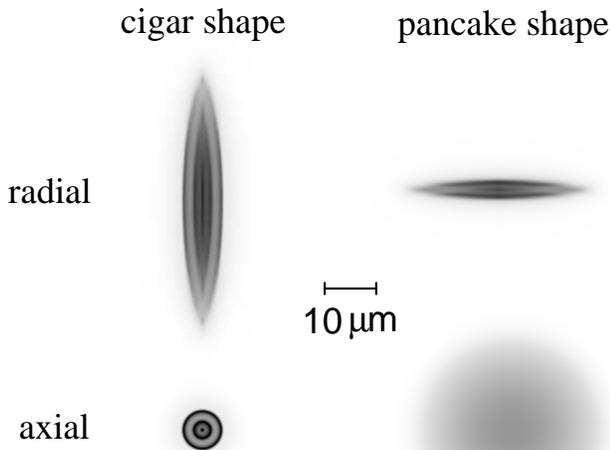}
\caption{
Pattern formation in a cigar-shape trap (left panel) and a pancake-shape
trap (right panel), where the column density is seen from the radial
(upper images) and axial (lower images) directions.
The initial number of atoms is $N_0 = 50000$, and at $t = 0$ the
s-wave scattering length is switched from $a_{\rm init} = 400 a_0$ to
$a_{\rm collapse} = -310 a_0$.
The ratio is $\omega_z / \omega_\perp = 0.39$ ($\sqrt{8}$) with
$\omega_\perp = 17.5$ Hz (9.03 Hz) in the cigar-shape (pancake-shape)
trap, and the images are taken at $t = 6.2$ ms ($t = 4.8$ ms).
}
\label{f:pattern}
\end{figure}
Here we predict that a similar pattern formation does occur in
${}^{85}{\rm Rb}$ BEC for experimentally available parameters.

Figure~\ref{f:pattern} shows pattern formation in axi-symmetric traps,
where $N_0 = 5 \times 10^4$, and the s-wave scattering length is switched
from $a_{\rm init} = 400 a_0$ to $a_{\rm collapse} = -310 a_0$.
The pancake-shape trap has the ratio $\omega_z / \omega_\perp = \sqrt{8}$
with the same geometric mean frequencies as the cigar-shape trap.
In the cigar-shape trap the cylindrical shell structure is formed, and in
the pancake-shape trap the layered structure is formed.

\begin{figure}[t]
\includegraphics[width=8.6cm]{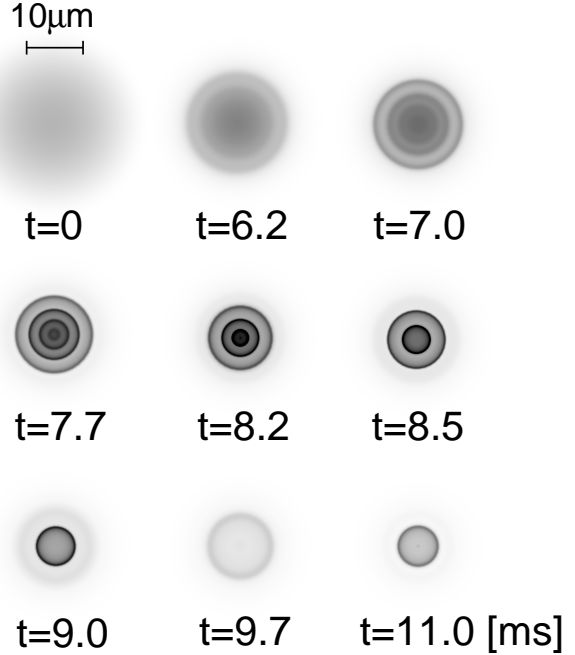}
\caption{
Time evolution of the column density of BEC in an isotropic trap, where
the trap frequency is $\omega / 2\pi = 12.8$ Hz.
At $t = 0$ the s-wave scattering length is switched from $a_{\rm init} =
400 a_0$ to $a_{\rm collapse} = -310 a_0$.
The initial number of atoms is $N_0 = 50000$ and the loss-rate
coefficient is $K_3 = 8 \times 10^{-26} {\rm cm}^6 / {\rm s}$.
}
\label{f:shells}
\end{figure}

Figure~\ref{f:shells} shows time evolution of the integrated column
density $\rho(x, z) = \int |\psi|^2 dy$ for an isotropic trap, where
$N_0$, $a_{\rm init}$, and $a_{\rm collapse}$ are the same as in
Fig.~\ref{f:pattern}.
We see that the density fluctuations grow to form four concentric
spherical shells at $t = 7.7$ ms, which move inwards and collapse one by
one; then at $t = 11$ ms we see that a new shell is being formed 

The resolution of the imaging system in Ref.~\cite{Donley} (7 $\mu{\rm m}$
FWHM) is inadequate for observing the patterns in Figs.~\ref{f:pattern}
and \ref{f:shells} (the spacing between the shells is $\sim 1 \; \mu {\rm
m}$).
Expansion of BEC before imaging will blur out the pattern.
In order to observe the pattern formation, therefore, we need to improve 
the {\it in situ} imaging method, or to use larger $d_0$ and $|a|$ to
enlarge the pattern.

\section{Summary}
\label{s:conclusion}

We have studied the dynamics of collapsing and exploding BECs by
numerically solving the time-dependent GP equation with atomic loss
(\ref{GP}), and compared our results with those of the experiments of
Ref.~\cite{Donley}.
We find that mean-field theory with atomic loss can account for the
following experimental findings:
(i) It takes the system a certain time $t_{\rm collapse}$ to undergo a
sudden decrease in the number of atoms after the jump of $a$.
(ii) The number of atoms in the condensate decays exponentially with a
decay time constant of a few milliseconds.
(iii) The burst atoms are ejected in the collapse process, and refocus
after every half trap period.
(iv) The fractions of remnant, burst, and missing atoms are almost
independent of $N_0$, and the number of remnant atoms is much larger than
the critical number $N_{\rm cr}$ for large $N_0$.
(v) The jets are observed when the collapse is interrupted by jumping $a$
to a positive value.

We have found that these phenomena are attributed to a rapid sequence of
local intermittent implosions, and provided a new interpretation of the
jets, i.e., the highly anisotropy of the jets is due to the interference
fringes.
This suggests that the burst atom cloud is coherent.

The validity of the mean-field GP equation is determined by the gas
parameter $n a^3$, and the depletion is given by $\simeq (n a^3)^{1/2}$.
When the implosion occurs, $n |a|^3$ becomes $\sim 10^{-3}$ at the peak
density in our simulations, which indicates that the mean-field
approximation is still valid at least qualitatively.

Our results presented here suggest that the mean-field approximation can
be used to describe the collapsing and exploding dynamics at least
qualitatively.
A more quantitative comparison between experiments and numerical
simulations might reveal effects beyond mean-field approximation.
This possibility merits further experimental and theoretical study.

\section*{ACKNOWLEDGMENTS}

We thank E. A. Donley for valuable comments.
This work was supported by a Grant-in-Aid for Scientific Research (Grant
No. 11216204) by the Ministry of Education, Science, Sports, and Culture
of Japan, and by the Toray Science Foundation.

\end{document}